\documentclass[usenatbib,usegraphicx]{mn2e}
\usepackage{aas_macros}
\usepackage{float}
\usepackage{times}
\newcommand{\swift}{J1753~}

\begin{document}

\title[420d modulation in Swift J1753.5-0127]{A 420 day X-ray/optical modulation and extended X-ray dips in the short-period transient Swift J1753.5-0127}
\author[A.W. Shaw et al.]{A. W. Shaw,$^1$ P. A. Charles,$^{1,2}$ A. J. Bird,$^1$ R. Cornelisse,$^{3,4}$ J. Casares,$^{3,4}$ F. Lewis,$^{5,6}$ \newauthor{T. Mu\~{n}oz-Darias,$^1$ D. M. Russell$^{3,4}$ and C. Zurita$^{3,4}$} \\
$^{1}$School of Physics and Astronomy, University of Southampton, Southampton SO17 1BJ, UK \\
$^{2}$Astrophysics, Cosmology and Gravity Centre (ACGC), University of Cape Town, Private Bag X3, Rondebosch, 7701, South Africa \\
$^{3}$Instituto de Astrof\'isica de Canarias (IAC), E-38200 La Laguna, Tenerife, Spain\\
$^{4}$Departamento de Astrof\'isica, Universidad de La Laguna (ULL), E-38206 La Laguna, Tenerife, Spain \\
$^{5}$Faulkes Telescope Project, University of Glamorgan, Pontypridd CF37 1DL, UK \\
$^{6}$Department of Physics and Astronomy, The Open University, Walton Hall, Milton Keynes MK7 6AA, UK}
\maketitle

\date{\today}
\begin{abstract}
We have discovered a $\sim420$d modulation, with associated X-ray dips, in RXTE-ASM/MAXI/Swift-BAT archival light-curves of the short-period (3.2h) black-hole X-ray transient, Swift J1753.5-0127.  This modulation only appeared at the end of a gradual re-brightening, approximately 3 years after the initial X-ray outburst in mid-2005. The same periodicity is present in both the 2-20 keV and 15-50 keV bands, but with a $\sim0.1$ phase offset ($\approx40$d).  Contemporaneous photometry in the optical and near-IR reveals a weaker modulation, but consistent with the X-ray period.  There are two substantial X-ray dips (very strong in the 15-50 keV band, weaker at lower energies) that are separated by an interval equal to the X-ray period. This likely indicates two physically separated emitting regions for the hard X-ray and lower energy emission. We interpret this periodicity as a property of the accretion disc, most likely a long-term precession, where the disc edge structure and X-ray irradiation is responsible for the hard X-ray dips and modulation, although we discuss other possible explanations, including Lense-Thirring precession in the inner disc region and spectral state variations. Such precession indicates a very high mass ratio LMXB, which even for a $\sim10M_\odot$ BH requires a brown dwarf donor ($\sim0.02M_\odot$), making Swift J1753.5-0127 a possible analogue of millisecond X-ray pulsars. We compare the properties of Swift J1753.5-0127 with other recently discovered short-period transients, which are now forming a separate population of high latitude BH transients located in the galactic halo.
\end{abstract}

\section{Introduction}
Soft X-ray transients, or X-ray novae (hereafter XRTs) are low-mass X-ray binaries (LMXBs) consisting of a neutron star (NS) or black hole (BH) compact object ($M_1$) accreting from a low-mass companion ($M_2$). $75\%$ of XRTs are believed to harbour a BH \citep{McClintock-2006}, and are characterised by long periods of quiescence (years to decades) followed by X-ray outbursts which can increase the luminosity by several orders of magnitude. Black hole X-ray transients (BHXRTs) have proven to be important in studying X-ray binaries (XRBs), as in quiescence they provide the opportunity to study the donor itself, which is mostly impossible in luminous, persistent XRBs \citep{Charles_Coe2006}.\\
\indent Swift J1753.5-0127 (hereafter J1753) was discovered by the Swift Burst Alert Telescope (BAT) in 2005 \citep{J1753-ATel} as a hard-spectrum ($\gamma$-ray source) XRT at high galactic latitude (+12$^{\circ}$). The source peaked within a week, at a flux of $\sim 200$ mCrab, as observed by the \textit{Rossi X-Ray Timing Explorer} (RXTE) All Sky Monitor (ASM) \citep{CadolleBel2007}. The source was also detected in the UV, with Swift's Ultraviolet/Optical Telescope (UVOT) \citep{Still-2005}, and in the radio with MERLIN \citep{Fender-ATel-2005}. An $R\sim15.8$ optical counterpart was identified by \citet{Halpern-2005}, who noted that it had increased by at least 5 magnitudes (as it is not visible on the DSS), thereby establishing \swift as an LMXB. Subsequent time-resolved photometry of the optical counterpart \citep{Zurita-2008} revealed R-band modulations on a period of $3.2$h, which are now well established.  By analogy with other XRTs (e.g. XTE J1118+480, see \citealp{Zurita-2002}), these were interpreted as a superhump period ($P_{sh}$), slightly longer than the orbital period ($P_{orb}$). \\
\begin{figure*}
\centering
\includegraphics[scale=0.8]{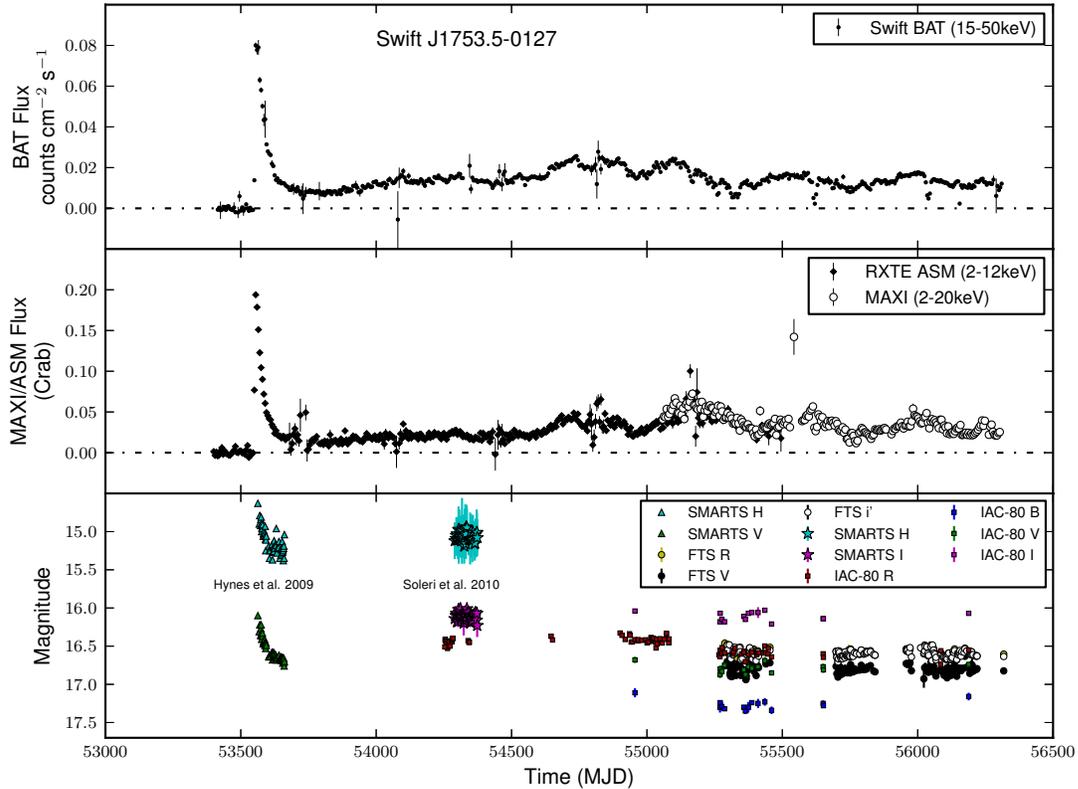}
\caption{\textit{Upper panel:} 15 - 50 keV Swift BAT light curve of \swift  spanning approximately 7 years from $30$ Jan $2005$ - $11$ Jan $2013$, using 5d binning. \textit{Centre panel:} Combined RXTE ASM (diamonds, 2-12 keV) and MAXI (open cirles, 2-20 keV) light curves, with both normalised to the Crab, again with 5d binning. \textit{Lower panel:} Combined optical and near-IR photometry of \swift from SMARTS (cyan (H) and green (V) triangles, \citep{Hynes-2009}; cyan (H) and magenta (I) stars, \citep{Soleri-2010}), IAC-80 (magenta (I), red (R), green (V) and blue (B) squares) and FTP (yellow (R), white \textit{i'} and black (V) circles).  In the upper and centre panels the dash-dotted line marks zero flux. A colour version of this figure is available in the online edition.}
\label{7yr_lc}
\end{figure*}
\indent Almost immediately after the outburst peak the X-ray flux of \swift started declining, but it then stalled at $\sim 20$ mCrab for over 6 months rather than returning to quiescence as might have been expected. Another well-known XRT to remain active for a significant period of time in this way is EXO 0748-676, a NS system and X-ray burster which remained active for 24 years before finally returning to quiescence \citep{Wolff-2008}. \swift unusually showed a steady increase in flux from late 2005, eventually producing an increase in hard band activity that was noted in mid 2008 \citep{Krimm2008-Atel} and at lower energies in late 2009 \citep{Negoro-2009}. Despite this long-term trend in X-ray flux, it remained in the low-hard (LH) state for $\sim4.5$ years after outburst, at which point it then underwent a complex transition to the hard intermediate state for a brief period before returning to the LH without passing through a soft state \citep{Soleri-2013}. \\
\indent With a large optical increase at outburst, it is not surprising that there has been no spectroscopic signature of the donor in \swift \citep{Durant-2009}, but with no detectable fluorescence features either, it has not been possible to obtain any direct indication of the compact object mass.  However, INTEGRAL observations highlighted the presence of a hard power-law tail up to $\sim 600$keV, very typical of a black hole candidate (BHC) in the hard state \citep{CadolleBel2007}.  Also the power density spectrum from a pointed RXTE observation revealed a $0.6$ Hz quasi-periodic oscillation (QPO) with a shape that is typically seen in BHCs \citep{Morgan-2005}. \swift is therefore a BHC with the third shortest $P_{orb}$ known (after Swift J1357.2-093313; \citealt{Corral-Santana-2013}, and MAXI J1659-152; \citealt{Kuulkers-2013}). Given the high galactic latitude of J1753, \citet{CadolleBel2007} concluded that its distance is likely $4-8$kpc, placing it in the galactic halo, similar to the BHXRT XTE J1118+480 \citep{Wagner-2001}.\\
\indent In this paper we present a detailed analysis of the long-term X-ray observations of J1753, using data from the RXTE ASM, Swift BAT and the \textit{Monitor of All-Sky X-ray Image} (MAXI) on-board the ISS \citep{Matsuoka-2009}, as well as monitoring from various optical telescopes. We focus here on the variability of the source over the near 7 year coverage provided by these facilities and in particular, note extended X-ray dips that are present in the BAT data.

\section{Observations and Analysis}
\subsection{Observations}
\indent \swift has been more or less continually observed by multiple X-ray to $\gamma$-ray instruments since its original outburst in 2005. Public data from MAXI, Swift BAT and the RXTE ASM have been used to produce the $\sim7$yr light curves presented here.  The data required no significant processing and are all available online. \footnote{http://maxi.riken.jp/top/} \footnote{http://heasarc.nasa.gov/docs/swift/results/transients/} \footnote{http://xte.mit.edu/} \\
\indent \swift  was also observed by the two 2m robotic Faulkes Telescope North (located at Haleakala on Maui) and Faulkes Telescope South (at Siding Spring, Australia), of the Faulkes Telescope Project (FTP), which are part of the Las Cumbres Observatory Global Telescope (LCOGT) network (e.g. \citealt{Faulkes-2010}).  Photometry was performed in the \textit{i'}, R and V-bands and the data were reduced using the FT pipeline. We also have optical data from the 80cm IAC-80 telescope at the Observatorio del Teide on Tenerife, and both the 1.5m and 0.84m telescopes at the Mexican Observatorio Astron\'{o}mico Nacional on San Pedro M\'{a}rtir \citep{Zurita-2008}.  \swift  has also been monitored in the H, I and V-bands with the Small and Moderate Aperture Research Telescope System (SMARTS) at Cerro Tololo \citep{Hynes-2009,Soleri-2010}.

\subsection{Light Curves}
\label{sec:lc}
\begin{figure}
\centering
\includegraphics[scale=0.4]{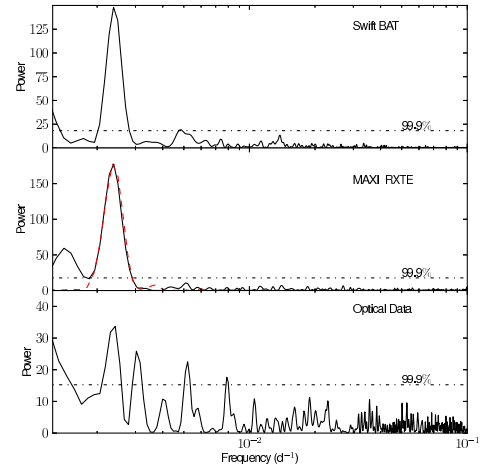}
\caption{\textit{Upper panel:} LS periodogram of the selected Swift BAT data of \swift (see text). \textit{Centre panel:} The solid line shows the periodogram of the combined RXTE ASM and MAXI light curves, whilst the red dashed line is a periodogram of a sine curve with the same period. \textit{Lower panel:} periodogram of the combined R and FTP \textit{i'}-band data of \swift from Fig. \ref{7yr_lc}. In all three plots the dash-dotted lines represent the 99.9\% confidence level. Both Swift BAT and MAXI/ASM power spectra exhibit rising power at very low frequencies; this is due to the gradual rise in flux during the data window. A colour version of this figure is available in the online edition.}
\label{periodograms}
\end{figure}

The 7 year X-ray light curve of \swift is shown in 5-day bins in Figure \ref{7yr_lc} and exhibits a profile typical of XRTs in outburst i.e. a FRED, or Fast Rise, Exponential Decay profile, with a peak at $\sim200$mCrab \citep{Ramadevi-2007}. However, after the initial outburst the flux stalled at $\sim20$mCrab for several months before it then started gradually increasing.  This behaviour is present in both the BAT and ASM light-curves, although we note that it is slightly steeper in the higher energy data.  Then, just after MJD $\sim54500$, there is a significant increase in flux, following which the light curve begins to exhibit a modulation with a period of $\sim400$d. The sudden increase in X-ray flux has been noted before \citep{Krimm2008-Atel,Negoro-2009}, but the long-term variability has not previously been investigated. \\
\indent We selected X-ray data for timing analysis by excluding the initial outburst/FRED portion, i.e. we used all data after MJD $54000$ (see Fig. \ref{7yr_lc}). The periods and their errors were determined using a Lomb-Scargle (LS) analysis to determine a central peak frequency, and then performing a Monte-Carlo simulation based on `bootstrap-with-replacement' to repeatedly extract a subset of the light curve to re-perform the periodogram analysis on. The distribution of peak frequencies resulting from 10000 iterations of this process was used to determine the error on the peak frequency. A LS power spectrum was also calculated using the available optical data in the combined R-band and FTP \textit{i'}-band data, as observations of the source in both these filters exhibited almost identical values. As seen in Figure \ref{periodograms}, there is a clear peak in both X-ray periodograms at $416.4\pm2.1$d (BAT) and $422.8\pm2.3$d (combined MAXI/ASM). The power spectrum of the combined optical data also exhibits a peak at a period of $419.2\pm3.5$d; however, whilst the peak power is still significant, it is lower than that in both X-ray power spectra, and the periodogram is noisier.  Nevertheless, it should be noted that these datasets are entirely independent of each other. \\
\indent When the light curves are folded on $P = 422.8$ days (Fig. \ref{folded}) it becomes clear that the peak of the hard X-ray light curve (Swift BAT) precedes that at lower energies (MAXI/ASM) by $\sim0.1-0.2$ phase ($\approx40-80$d). There is also evidence of potential dipping structure in the folded Swift BAT light curve, which we explore further in \S\ref{sec:dips}. The folded optical light curve shows evidence of a tentative anticorrelation between the optical and the hard X-ray flux, with a correlation coefficient $r\approx-0.3$ (at 95\% confidence), with a similar value of $r$ being calculated for the optical vs. MAXI data. The confidence level was determined from the two-tailed p-value calculated alongside the Pearson $r$ coefficient.

\subsection{X-ray dips}
\label{sec:dips}
\indent One of the most remarkable features of the Swift BAT light curve of \swift is the presence since MJD $\sim55500$ of two extended X-ray dips, which are separated by 420 days, an interval that is fully consistent with the long term modulation discussed in \S \ref{sec:lc}. These dips last for $\approx25$ days and are almost a full eclipse of the hard X-ray flux, and occur at the maximum of the MAXI/ASM modulation. The dips in the Swift BAT light curve at MJD $\sim55625$ and $\sim56045$ are clearer in the more detailed Fig. \ref{dips}, where related, but weaker, features in the lower energy MAXI light curve can also be seen. The MAXI feature is more significant in the first dip ($\sim30$ mCrab) than in the second. Dips on this timescale are not visible anywhere else in the near 7 year light curves of the source in either energy band, which is why they are not immediately obvious in the folded Swift BAT light curve in Fig. \ref{folded}. Figure \ref{dips} also includes contemporaneous optical monitoring data that has been obtained with FTP and IAC-80. Evidence of optical dips in these data is marginal, although there may be a tentative dip of $\sim0.15$ magnitude, coincident with the second hard X-ray dip, in the FTP R, V and \textit{i'}-band data. However, this variation is within the broad range of fluctuations seen on long timescales, so cannot be considered significant.

\section{Discussion}
\subsection{The 420d modulation}
\indent \swift  is a unique candidate BHXRT in that the source has not returned to quiescence like most transients, but instead is now exhibiting a substantial long-term modulation over a wide range of energies on a period of $\sim420$d.  Since we know $P_{orb}\approx3.2$h \citep{Zurita-2008}, this modulation is a ``super-orbital'' periodicity of the kind that has now been seen in a significant fraction of XRBs.  There are a variety of mechanisms that can give rise to long-term (tens to thousands of days) variations in all types of XRBs (see e.g. \citealt{Charles-2010,Kotze-2012}), but the most likely cause in this case is disc precession as a result of the high mass ratio in BH LMXBs.  Such behaviour is well-established in the CV analogues of LMXBs, the SUMa systems, which exhibit ``superhumps'' on periods very slightly longer than $P_{orb}$ (see e.g. \citealt{Warner-1995}), as a result of the disc expanding outside its tidal stability radius (for full details see e.g. \citealt{Whitehurst-1991}).  More importantly, \citet{Patterson-2005} have demonstrated an observational link between the period difference (i.e. $P_{sh}-P_{orb}$) and mass ratio $q=M_2/M_1$, and this relationship has been extended from CVs to LMXBs.  Indeed, it was applied by \citet{Zurita-2002} in their analysis of XTE J1118+480. \\
\begin{figure}
\centering
\includegraphics[scale=0.4]{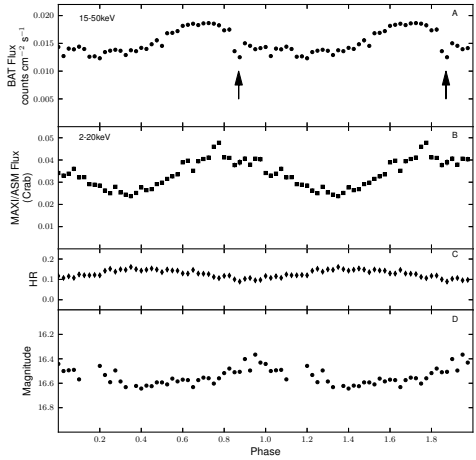}
\caption{\textit{Panel A:} Swift BAT light curve of \swift  folded on a period of $422.8$ days (as determined from the MAXI/ASM periodogram in Fig. \ref{periodograms}), with arrows marking the location of hard X-ray dips that are discussed in the text. \textit{Panel B:} Combined MAXI \& ASM light curve of \swift folded on the same period. \textit{Panel C:} Hardness Ratio (HR) defined as BAT count rate/MAXI \& ASM count rate. \textit{Panel D:} Combined R and Faulkes \textit{i'}-band light curves of \swift folded on the same period. In all panels, each bin represents $\approx$ 10 days.}
\label{folded}
\end{figure}
\indent Consequently, the presence of such long-term timescales in BHXRTs is taken as evidence of a precessing accretion disc, and hence can be extremely important in being able to provide an indication of $q$ without any direct kinematic measurements.  \citet{Zurita-2008} claimed a tentative precession period of $29$d from their R-band photometry. This can be interpreted as the beat between the orbital and superhump frequencies i.e. $P_{prec} = (P^{-1}_{orb}-P^{-1}_{sh})^{-1}$, giving $P_{orb}=3.23$h for the observed $P_{sh}=3.2443$h. This would imply $q=0.025$, using $\Delta P= (P_{sh} - P_{orb} / P_{orb}) = 0.18q + 0.29q^2$ \citep{Patterson-2005}. If we instead re-apply the Patterson relation using our $\sim420$d modulation as $P_{prec}$, we find that this gives a much more extreme $q\sim0.002$.  If the compact object is a typical $\sim10M_{\odot}$ black hole \citep{Farr-2011}, then the donor has a mass of $\sim0.02 M_{\odot}$, and hence is itself highly evolved.  \\
\indent Such an extreme $q$ suggests a comparison of \swift with SAX J1808.4-3658, the first accreting millisecond X-ray pulsar (AMXP), with $P_{orb}=2$h \citep{Chakrabarty-1998}. With an estimated donor mass $M_2 \approx 0.05 M_{\odot}$ \citep{Bildsten-2001}, we suggest that \swift is a BH analogue to this AMXP system, taking the mass ratio to new extremes.  If this is the case, a $0.02 M_{\odot}$ brown dwarf (BD) donor would require a radius of $\approx0.14 R_{\odot}$ to sustain Roche Lobe overflow, and this is consistent with the M-R relation of low-mass stars and sub-stellar objects detailed in Fig. 2 of \citet{Chabrier-2009}. This relation is approximately constant below masses of $\sim0.1 M_{\odot}$, which may account for the extended duration of the \swift outburst, as the BD radius is roughly independent of mass.\\
\indent Another mechanism we considered for the long-term modulation of \swift was that of irradiation-driven warping. In this case an initially flat accretion disc is unstable to warping when irradiated by a central source, as the warp then excites the tilt of the disc continuously, thereby leading to precession \citep{Ogilvie-2001, Foulkes-2006}. This model was used to simulate real binary systems with observed or inferred super-orbital periods by \citet{Foulkes-2010}, who found they could account for observed X-ray luminosities for a given super-orbital period ($P_{sup}$). However, high $q$ systems were found to have the shortest $P_{sup}$ out of all XRBs examined, and knowing that the donor of \swift must be very low mass (it isn't visible on the DSS), it is unlikely that this scenario applies to J1753.\\

\subsection{Properties of the 420d modulation}
\indent The folded light curves in Fig. \ref{folded} show that the hard X-ray emission leads that at lower energies, but the hardness ratio (HR) modulation (Fig. \ref{folded}, panel C) shows that this could also be interpreted as a smooth spectral variation. This is reminiscent of that seen in 4U 1636-536, an LMXB with a $3.8$h period that displays long-term variability on timescales of $30-40$d \citep{Belloni-2007}. Remarkably, while the long-term variability was found at both soft ($2-12$ keV, RXTE ASM) and hard ($20-100$ keV, INTEGRAL IBIS) X-ray energies, it was also found to be anticorrelated \citep{Shih-2005}. The long-term variation was initially suggested to be due to a tilted or warped precessing accretion disc \citep{Shih-2005}, and the measured mass ratio allowed for this hypothesis \citep{Casares-2006}.  However, the variations were not stable and coupled with the anticorrelation between soft and hard X-rays, it is difficult to interpret as a precessing disc \citep{Shih-2011}. Regardless of the nature of the long-term variations, the soft/hard X-ray anticorrelation seen in the 4U 1636-536  light curves suggests physically separate emission regions responsible for hard and soft X-rays, a hypothesis we believe can be applied to J1753. This theory is further strengthened by the presence of strong X-ray dips in the Swift BAT light curve of J1753, but only weak dips in the softer MAXI light curve.

\subsection{Hard X-ray dips}
\begin{figure*}
\centering
\includegraphics[scale=0.8]{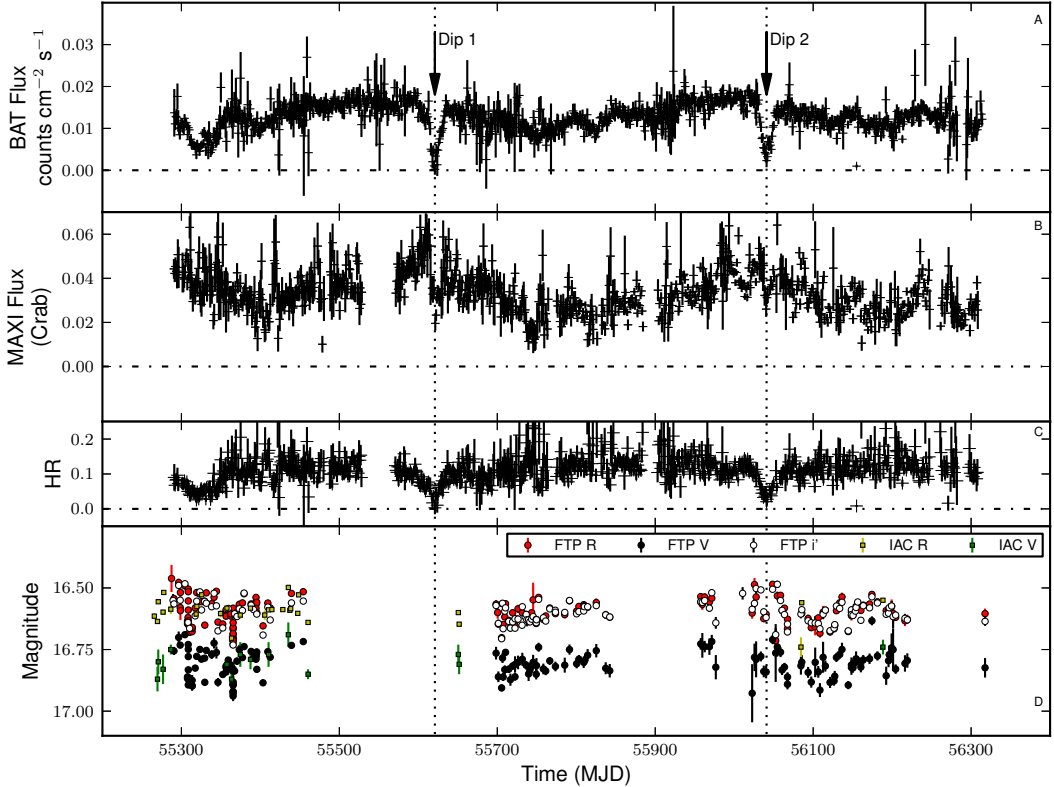}
\caption{\textit{Panel A:} Swift BAT light curve (15 - 50 keV) of \swift with 1d binning. \textit{Panel B:} 2 - 20 keV light curve of \swift  from MAXI, also in 1d bins. \textit{Panel C:} Hardness Ratio (HR) defined as BAT count rate/MAXI count rate. \textit{Panel D:} Optical photometry of \swift from the IAC-80 (yellow (R) and green (V) squares) and FTP (red (R), black (V) and white (\textit{i'}) circles). In panels A, B and C, the dash-dotted line marks the zero point and the vertical dotted lines highlights the minimum of the dips. A colour version of this figure is available in the online edition.}
\label{dips}
\end{figure*}

\indent The hard X-ray dips could be attributed to a warped disc structure that obscures the hard X-ray emitting region entirely whilst leaving the softer region mostly visible. We presume that the dips do not appear earlier in the overall light curve as the warp was not yet established at that time. The dip light curves can be used to place constraints on the size of the hard X-ray emitting region if we adopt our derived value of $q$. Assuming that the obscuring material is located at the edge of the disc ($R_1\approx1.8R_{\odot}$), the angle swept out by the warp would be $\sim21^{\circ}$, from which we roughly estimate a hard X-ray emitting region size of $\sim0.3$ $R_{\odot}$. However, we note that the lack of similar scale dips at lower energies implies that the latter is from a significantly more extended region. \\
\indent We also note that the dips are prominent in the HR light curve (Fig. \ref{dips}, panel C), which shows similar behaviour in the more extended, but shallower dip at MJD $\sim55300$. Interestingly this is the interval when \citet{Soleri-2013} observed spectral state changes, and so these dips could all be related to such behaviour, although it is not clear why this would be periodic.

\indent Finally we note that the 420d separation of the two dips may be consistent with Lense-Thirring precession at the Bardeen-Petterson radius \citep{Bardeen-1975}. It is at this radius that frame dragging effects from the rapidly rotating black hole can give rise to twisted, warped structure in the inner disc, which has been linked to the QPOs seen in many XRBs \citep{Fragile-2001}, and could offer an explanation for the hard X-ray dips. Linking the 0.6Hz QPO seen in \swift with the Bardeen-Petterson radius would place it at $\sim200$ $R_s$ (where $R_s$ is the Schwarzschild radius of the BH) for a typical BH mass, which could have a Lense-Thirring precession period comparable to our observed 420d. However, these values are highly dependent on the scaling and spin parameters of the black hole as described in Equation 4 of \citet{Fragile-2001} and such effects have never been seen at these timescales.

\subsection{Comparison with other high latitude X-ray transients}
\begin{table*}
\caption{Black-Hole X-Ray Transients in the Galactic Halo} \label{table}
\centering
    \begin{tabular}{|p{3cm} c c c c c c c c c c|}
    \hline
    Source & $F_x$ $^a$ & $b$ & $d$ & $v_D$ $^b$ & $P_{orb}$ & $P_{sup}$&q&$M_1$ & \textit{i}& $M_2$\\
      & (mCrab) & & (kpc) & ($km$ $s^{-1}$) & (hr) & (d) & & ($M_{\odot}$) & & ($M_{\odot}$) \\
    \hline
    GRO J0422+32$^1$ & 3000 & $-12^{\circ}$ & $\sim2.5$ & 400 & $5.1$ & ?& 0.075 &$\sim4$ & $45^{\circ}$ & $0.3$ \\
    XTE J1118+480$^2$ & $\sim40$ & $+62^{\circ}$ & $\sim1.8$ & 600 & $4.1$& $\sim52$ & $< 0.025$ & $8$ & $68^{\circ}$ & $0.2$ \\
    MAXI J1305-704$^3$ & $\sim30$ & $-7^{\circ}$ & $?$ & 500*& ? & ? & $?$ & $?$ & $?$ & $?$ \\
    Swift J1357.2-093313$^4$ & $\sim30$& $+50^{\circ}$ & $\sim1.6$ & 900 & 2.8& ? &$< 0.06$ & $>3$ & $\geq70^{\circ}$ & $0.2$\\
    MAXI J1659-152$^5$ & $\sim50$& $+16^{\circ}$ & $8.6$ & ? & 2.4 & ? & $< 0.08$ & $>3$ & $\sim65^{\circ}-80^{\circ}$ & $0.15-0.25$\\
    Swift J1753.5-0127 & $\sim200$& $+12^{\circ}$ & $\sim8$ & 600 & 3.2 & $\sim420$ & $\sim0.002$ & $\sim10$ & $?$ & $\sim0.02$\\
    \hline
    \end{tabular}
\small \raggedright \\ Note: $^a$ Peak X-ray flux, $^b$ Disk velocity estimated from $H_{\alpha}$ double peak separation \citep{Warner-1995}.\\
\** denotes that $v_D$ was estimated from HeII emission. \\ 
$^1$\citealt{Gelino-2003}, $^2$\citealt{Zurita-2002}, $^3$\citealt{Charles-2012}, $^4$\citealt{Corral-Santana-2013}, $^5$\citealt{Kuulkers-2013} \\
\end{table*}

Remarkably the last decade has seen the emergence of a number of short period BHXRTs which we can compare with J1753. Interestingly these are all at high Galactic latitude and hence located in the halo (Table \ref{table}), and may represent a sub-class of BHXRTs which are difficult to find in soft X-ray surveys, but are seen by Swift BAT in the LH state. \\
\indent It is interesting to note that 4 of the 6 sources listed show dipping structure. Swift J1357.2-093313 exhibits very fast optical dips ($\sim$2-8min) and these cannot be Keplerian at the outer disc, but must be in the inner disc region \citep{Corral-Santana-2013}. Swift J1357.2-093313 shows no X-ray dips and thus must have a very high $i$, requiring that the central BH is hidden from view completely, and thus we only see scattered X-rays, accounting for the low peak $F_x\sim30$mCrab, making such systems hard to detect. \\
\indent On the other hand, MAXI J1659-152 exhibits X-ray dips suggestive of a disc edge structure, hence the range in $i$ shown in Table \ref{table}. This highlights the importance of considering the geometry of such systems, as we would not expect this high a fraction of dipping sources within the sample.

\subsection{Future Work}
\indent If the hard X-ray dips are indeed a periodic phenomenon, then we predict that the next dip will occur in June 2013, with the minimum being reached $\sim$Jun 18. We will monitor both the Swift BAT and the MAXI light curves during this time. We also intend to obtain pointed observations with both optical and X-ray telescopes in order to determine the structure of the dips at multiple wavelengths. An ideal instrument to cover the dip structure would be Swift, as a pointed observation would allow simultaneous use of the X-ray Telescope (Swift XRT) and UVOT, as well as the continued hard X-ray monitoring with BAT. Contemporaneous soft and hard X-ray and UV/Optical monitoring will allow us to provide constraints on the nature of the dips.

\section*{Acknowledgements}
We would like to thank the anonymous referee for the helpful comments provided. RC acknowledges a Ramon y Cajal fellowship (RYC-2007-01046). RC and JC acknowledge support by the Spanish Ministry of Science and Innovation (MICINN) under the grant AYA 2010-18080. DMR acknowledges support from a Marie Curie Intra European Fellowship within the 7th European Community Framework Programme (FP7) under contract no. IEF 274805. TMD
 acknowledges funding via an EU Marie Curie Intra-European Fellowship under contract no. 2011-301355. The Faulkes Telescopes (North and South) are maintained and operated by LCOGT. This research has made use of MAXI data provided by RIKEN, JAXA and the MAXI team. We acknowledge the use of public data from the Swift data archive.

\footnotesize{
\bibliography{references.bib}
}
\end{document}